\documentclass[journal]{IEEEtran}

\usepackage{stfloats}
\usepackage{amssymb,amsmath}
\usepackage[pdftex]{graphicx}
\usepackage{caption}
\usepackage{mathtools}
\usepackage{enumitem}
\usepackage[font=small]{caption}
\usepackage{array}
\usepackage{multicol}
\usepackage{blindtext}
\usepackage{bm}
\usepackage{mathrsfs}

\usepackage{cite}
\usepackage[pdftex]{graphicx}
\usepackage{amssymb,amsmath}
\usepackage{float}
\usepackage{bm}
\usepackage{mathtools}
\usepackage{amsmath}
\usepackage{array}
\usepackage{multirow}
\usepackage{tabularx,booktabs}
\usepackage{stfloats}
\usepackage[utf8]{inputenc}
\usepackage[english]{babel}
\usepackage{hyperref}
\usepackage{array}
\usepackage{subcaption}
\usepackage{bm}
\usepackage{etoolbox}
\AtBeginEnvironment{algorithm}{\setstretch{2}}
\usepackage{setspace}
\newcolumntype{L}[1]{>{\raggedright\let\newline\\\arraybackslash\hspace{0pt}}m{#1}}
\newcolumntype{C}[1]{>{\centering\let\newline\\\arraybackslash\hspace{0pt}}m{#1}}
\newcolumntype{R}[1]{>{\raggedleft\let\newline\\\arraybackslash\hspace{0pt}}m{#1}}

\usepackage{listings}
\usepackage{color} 
\definecolor{mygreen}{RGB}{28,172,0} 
\definecolor{mylilas}{RGB}{170,55,241}

\usepackage{algorithm,algpseudocode}
\makeatletter
\renewcommand{\Function}[2]{%
  \csname ALG@cmd@\ALG@L @Function\endcsname{#1}{#2}%
  \def\jayden@currentfunction{#1}%
}
\newcommand{\funclabel}[1]{%
  \@bsphack
  \protected@write\@auxout{}{%
    \string\newlabel{#1}{{\jayden@currentfunction}{\thepage}}%
  }%
  \@esphack
}
\makeatother
\makeatletter
\newcommand{\linebreakand}{%
  \end{@IEEEauthorhalign}
  \hfill\mbox{}\par
  \mbox{}\hfill\begin{@IEEEauthorhalign}
}
\makeatother

\begin{document}

\title{Quantifying Implicit Overload Mandates in Phase Jump Requirements for Grid Forming Inverters}

\author{
\IEEEauthorblockN{M~A Awal, Rahul Chakraborty, David Michaud, Mikko Qvintus, Devin Dilley}

\thanks{M~A~Awal, David~Michaud, Mikko~Qvintus, and Devin~Dilley with EPC Power Corporation, Poway, CA, USA (e-mail: m.awal@epcpower.com; david.michaud@epcpower.com, mikko.qvintus@epcpower.com, bill.giewont@epcpower.com, devin.dilley@epcpower.com).}

\thanks{Rahul~Chakraborty is with Dominion Energy, VA, USA (e-mail: rahul.chakraborty@dominionenergy.com).}
}
\maketitle

\begin{abstract}
Grid codes increasingly require grid-forming (GFM) inverters to demonstrate prescribed active-power response to phase-angle jumps at the point of interconnection (POI). This paper shows that such requirements embed an implicit current-overload mandate whose severity depends on the test parameters but is nowhere made explicit in the specifications. First, an analytic expression for the instantaneous power is derived at an arbitrary measurement point, establishing that a momentary power excursion in the non-opposing direction is an inevitable physical consequence of the phase jump itself, independent of control action. Second, the phase-jump recovery is formulated as a constrained optimal control problem with the characteristic GFM objective of minimizing terminal voltage deviation from the pre-disturbance value while subject to a hard current limit. As the plant dynamics are linear and the constraints are convex, the solution constitutes a controller-architecture-independent physical bound on the achievable power-recovery trajectory. Sweeping the current limit, the phase-jump acceptance criterion is converted into an equivalent minimum overload ratio, making the implicit hardware mandate quantitative. The bound is validated against three WECC generic GFM inverter models (REGFM\_A1, B1, C1) in electromagnetic transient simulations, confirming both validity and tightness of the bound. Recommendations are offered for interpreting compliance test results and for structuring test specifications to distinguish physical hardware limitations from control deficiencies.
\end{abstract}
\vspace{10pt}


\IEEEpeerreviewmaketitle

\bstctlcite{IEEEexample:BSTcontrol}

\section{Introduction}
\label{sec:introduction}
The displacement of synchronous generation by inverter-based resources (IBRs) is fundamentally reshaping the dynamic characteristics of bulk power systems. As IBR penetration grows, system operators confront the loss of attributes that synchronous machines provided intrinsically, such as rotational inertia, fault current contribution, and a self-synchronizing voltage-source interface with the grid \cite{nerc_ibr_strategy,matevosyan2019grid}. Grid-forming (GFM) inverter control has emerged as the leading candidate to restore these properties, and several jurisdictions have begun codifying GFM requirements into interconnection standards \cite{ieee2800,ercot_ags_esr, aemo_gfm,nerc_gfm_wp,ngeso_gbgf}.

A grid-forming inverter is characterized by its \emph{voltage source behind reactance} behavior, where the voltage vector's magnitude and frequency are dynamically regulated locally following droop laws, enabling the device to respond immediately and autonomously to changes in the external grid \cite{aemo_gfm,rosso2021review}. This voltage-source behavior stands in contrast to grid-following control, which regulates current injections in response to a measured grid voltage \cite{rosso2021review,du2020comparative}. GFM control structures, such as droop control, 
virtual synchronous machine (VSM), and virtual oscillator control (VOC), all share the same power-synchronization principle but differ in transient response and implementation details \cite{darco2014equivalence,zhong2011synchronverters,johnson2016synthesizing,lasseter2020grid}.

Among the suite of compliance tests now being developed for GFM inverters, the phase-angle-jump test occupies a unique position. Unlike frequency ramps or symmetric voltage steps, a phase jump at the infinite bus alters the angle between the inverter's internal voltage vector and the grid voltage instantaneously, probing the inverter's ability to maintain synchronism and deliver power through a large-signal transient \cite{ercot_ags_esr,aemo_gfm}.  Majority of emerging grid codes, such as \emph{ERCOT Advanced Grid Support Energy Storage Resource (AGS-ESR) functional 
specification}, \emph{AEMO Voluntray Specification for Grid Forming Inverters}, \emph{NESO Minimum Specification Required for Provision of GB Grid Forming Capability}, mandate some form of the \emph{Phase-Jump Test} with variations in the test conditions. The distinctions among the test conditions arise from the short-circuit ratio (SCR), the $X/R$ ratio, pre-disturbance operating point, maximum phase jump amount as well as the performance 
criteria. For instance, ERCOT specification uses a pre-disturbance operating condition of 1pu real power injection by a GFM BESS and requires that for a phase jump $\Delta \theta \in \bigl\{\pm 10^\circ, \pm 25^\circ\bigr\}$, the active-power must change to oppose the phase jump and the peak change from the pre-disturbance active power level must exceed 0.2\,pu per 10$^\circ$. Furthermore, the rise time to 90\% of the initial change needs to be within one cycle, and the active power cannot not return to its pre-disturbance level for at least three cycles \cite{ercot_ags_esr,ercot_ags_faq}. Notably, when the inverter reaches its current limit, the specification acknowledges that these criteria ``may not be applicable,'' but still requires power to ``return to the pre-disturbance level in a stable manner'' \cite{ercot_ags_faq}. In contrast, AEMO specifies the pre-disturbance operating point as 0.5pu real power injection and phase-jump steps $\Delta \theta \in \bigl\{\pm 10^\circ, \pm 30^\circ, \pm 60^\circ \bigr\}$ with similar relaxation provision under current limit. The relaxation provisions with somewhat ambiguous performance mandates leave substantial room for interpretation and reveal a structural gap in the specification: the phase-jump acceptance criteria implicitly demand a transient current capability that 
exceeds the continuous rating, yet no explicit overload ratio is specified.
\begin{figure}[htb]
	\makebox[\linewidth][c]{\includegraphics[angle = 0, clip, trim=0cm 0cm 0cm 0cm, width=0.5\textwidth]{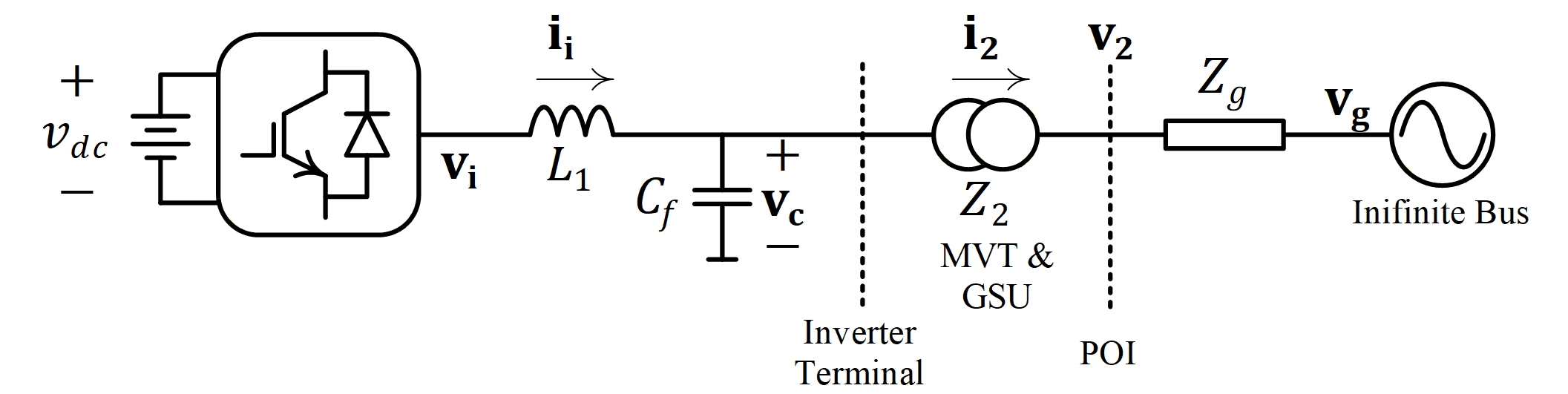}}
	\caption{Test configuration of interest.}
	\label{fig:topology}
    \vspace{-10pt}
\end{figure}
\subsection{Topology}
To contextualize the structural gap, consider a GFM inverter, with current overload rating of 1.25pu, dispatched at rated power ($P_{pre} = 1$pu, $Q_{pre} = 0$pu) when a negative phase jump $\Delta \theta = -10^\circ$ occurs. The ERCOT AGS-ESR specifications provide explicit performance criterion in this case requiring the ESR to provide a monotonic real power excursion above the pre-disturbance level reaching a peak of 1.2pu. However, for a phase jump of $\Delta \theta = -25^\circ$, the ESR hits its current limit before reaching the desired 1.5pu real power peak, which allows the relaxation provision---only requiring ``return to pre-disturbance level in a stable manner." The lack of explicit performance specification in such a scenario often leads to disagreements among different parties during the interconnection approval process. In particular, one may interpret the requirement as a monotonic rise in real-power until the ESR reaches its current limit; however, such an interpretation, although reasonable prima facie, fails to recognize a hard physical bound. This work contends that monotonic rise in real power under such condition mandates a higher transient over-current capability, and the required overload ratio is quantified through rigorous analysis for given test conditions such as the phase jump magnitude and the grid strength (SCR)---none of which are tied together quantitatively in the existing specifications.

The interaction between current limiting and GFM transient stability has received significant attention in recent literature. Rokrok \emph{et al.}\ analyzed the impact of the current-reference saturation angle on the critical clearing time, demonstrating that the choice of limiting strategy can influence transient synchronization stability \cite{rokrok2022transient}. Qoria \emph{et al.}\ compared virtual-impedance and current-saturation algorithms, showing that virtual impedance improves transient stability margins at the cost of initial oscillations \cite{qoria2020current,qoria2020virtual}. Taul \emph{et al.}\ proposed an enhanced current-limiting scheme that maintains GFM dynamics during faults \cite{taul2020current}. Kkuni and Yang presented a quantitative analysis of the transient stability margins, including maximum survivable phase jump, under current-limiter operation 
for droop-controlled converters \cite{kkuni2024effects}. These studies share a common focus on \emph{whether} the inverter remains stable under current saturation, but do not address the converse question central to this paper: \emph{what minimum overload capability is physically required} to meet a given set of phase-jump performance criteria?

This paper addresses that question by formulating the phase-jump recovery as a constrained optimal control problem (OCP) on the physical inverter hardware including the input filter and DC voltage bound along with the AC grid conditions.
The key contributions are as follows:
\begin{enumerate}
  \item A controller-architecture-independent physical bound on the achievable voltage-tracking and power-recovery trajectory is established by solving the constrained OCP offline. As the plant dynamics are linear and the current and modulator constraints are convex, the solution represents the best any realizable GFM control can achieve for a given hardware configuration.
  
  \item An analytic expression for the instantaneous power at an arbitrary measurement point, such as the POI in a utility-scale ESR plant, showing that a non-monotonic real power excursion, specifically a momentary drop, in response to a negative phase jump is inevitable.
  
  \item By sweeping the current limit against power-excursion targets, any phase-jump acceptance criterion is converted into a minimum overload ratio, making the implicit hardware mandate quantitative and transparent.
  
  \item The physical bound is substantiated against predominant GFM control architectures through electromagnetic transient (EMT) simulations, quantifying the performance gap between each architecture and the theoretical frontier.
\end{enumerate}

The remainder of this paper is organized as follows. Section~II presents the system model and the dq-frame state-space formulation. Section~III formulates the constrained OCP with the GFM voltage-tracking objective. Section~IV derives the measurement-point power expression and the geometric initial-condition analysis. Section~V presents the optimal trajectories and the current-limit sensitivity sweep. Section~VI compares the physical bound against specific GFM controller architectures. Section~VII discusses the implications for grid-code development, and Section~VIII concludes the paper.

\section{System Model}
In a typical battery energy storage system (BESS) plant rated at hundreds of MWs, a multitude of low voltage (480V AC - 690V AC) inverters, each rated hundreds of kW to few MW, are used with their respective battery stacks. Dedicated medium voltage transformers (MVTs) step-up the LV to the medium voltage levels, such as 13.8kV AC--34.5kV AC. Finally, the MV voltage distribution grid in the plant eventually connects to the sub-transmission or transmission level, such as 230kVAC, through a generation step-up transformer (GSU). Fig.~\ref{fig:topology} depicts an equivalent system model, where an aggregate model is used and the MVTs and GSU are lumped together; an LC type input filter is used in the inverter. Such lumped/aggregate models are widely used in compliance testing during the interconnection approval process.

In the following analysis all quantities are expressed in per-unit on the aggregate plant's rated capability $P_b$ and voltage base $V_b$. Inductances and capacitances are given as per-unit reactance and susceptance, respectively, at the nominal angular frequency $\omega_b = 2\pi f_b$: $L_{\mathrm{pu}} = (\omega_b L)/Z_b,\ C_{\mathrm{pu}} = \omega_b C Z_b$, where $Z_b = V_b^2 / S_b$. Time remains in seconds throughout. The inverter's input filter is modeled by $L_1$ and $R_1$ denoting the inductance and equivalent series resistance (ESR) of the input filter choke, 
$C_f$ and $R_f$ denoting the capacitance and ESR of the Y-connected shunt capacitors, respectively. The grid-side chokes in the input filter are typically small or may even be omitted relying on the MVT and hence we treat the shunt capacitor node as the inverter terminal without loss of generality. The MVT and GSU are modeled by lumped parameters $L_2$ and $R_2$, whereas the test grid is modeled by an \emph{infinite-bus} voltage source $\mathbf{v_g}$ and the line impedance $Z_g$ defined as $Z_g = 1/\mathrm{SCR}$ and $R_g = L_g/(X\!/\!R)$. For compliance test, various perturbations are introduced in the magnitude, frequency, and phase of infinite-bus voltage source $\mathbf{v_g}$ in electromagnetic transient (EMT) simulation environment using manufacturer provided inverter models, which are benchmarked against original hardware performance under various steady-state and transient conditions. Performance is measured at the POI and quantified against the specifications mandated by the relevant grid regulator.

\subsection{State-Space Model}
We adopt a synchronous reference frame rotating at the nominal grid frequency $\omega_0=\omega_b$, whose 
$d$-axis is aligned with the infinite-bus voltage $\mathbf{v_g}$. The state vector, control input, and 
grid-voltage input are given by
\begin{equation}
  \begin{aligned}
    \mathbf{x} = &\begin{bmatrix} \mathbf{i}_1^\top & \mathbf{v}_c^\top & \mathbf{i}_2^\top\end{bmatrix}\!^\top;\quad \mathbf{u}=\mathbf{v}_i; \quad \mathbf{w}=\mathbf{v}_g,\\
  \end{aligned}
  \label{eq:state_def}
\end{equation}
\noindent
where $\mathbf{i}_1=[i_{1d}\ i_{1q}]^\top$, $\mathbf{i}_2=[i_{2d}\ i_{2q}]^\top$, $\mathbf{v}_c=[v_{cd}\ v_{cq}]^\top$, and $\mathbf{v}_g=[v_{gd}\ v_{gq}]^\top$. The state dynamics is derived as
\begin{align}
  \dot{\mathbf{i}}_1 &= \omega_b/L_1\cdot \bigl[\mathbf{v}_i - \mathbf{v}_c - (R_1+R_f) \cdot\mathbf{i}_1\bigr] + \mathbf{\Omega}\cdot \mathbf{i}_1  \label{eq:di1} \\
  \dot{\mathbf{v}}_c &= \omega_b/C_f\cdot \bigl(\mathbf{i}_1 - \mathbf{i}_2\bigr) + \mathbf{\Omega}\cdot \mathbf{v}_c  \label{eq:dvc} \\
  \dot{\mathbf{i}}_2 &= \omega_b/L_t\cdot \bigl[\mathbf{v}_c - \mathbf{v}_g - (R_t-R_f) \cdot\mathbf{i}_2\bigr] + \mathbf{\Omega}\cdot \mathbf{i}_2  \label{eq:di2}
\end{align}
\noindent
where $L_t=L_2+L_g$, $R_t=R_2+R_g$, and $\mathbf{\Omega}=\begin{bmatrix}0 & \omega_0\\-\omega_0 & 0\end{bmatrix}$. The state dynamics is compactly written as
\begin{equation}
  \dot{\mathbf{x}} = A\,\mathbf{x} + B\,\mathbf{u} + B_g\,\mathbf{w},
  \label{eq:ss}
\end{equation}
\noindent
which represents a \emph{linear and time-invariant (LTI)} dynamics, whereas the converter voltage $\mathbf{u}$ enters through $B$, and the grid voltage $\mathbf{w}$ enters through $B_g$. All nonlinearity in the physical problem resides in the \emph{constraints}, not the dynamics.

\subsection{Measurement at POI}
\label{sec:vmeas}
The output current $\mathbf{i_2}=[i_{2d}\ i_{2q}]^\top$ at the POI is readily available from the state dynamics given by \eqref{eq:ss}, the POI voltage $\mathbf{v_2}=[v_{2d}\ v_{2q}]^\top$, at the intersection of $Z_2$ and $Z_g$, is 
derived algebraically following Kirchhoff's voltage law and eliminating $\dot{\mathbf{i}}_2$ via \eqref{eq:di2}, yielding,
\begin{equation}
  \mathbf{v}_2 = L_g/L_t\cdot \mathbf{v}_c + L_2/L_t\cdot \mathbf{v}_g + (L_2 R_g - L_g R_2)/L_t\cdot \mathbf{i}_{2}
  \label{eq:vmeas}
\end{equation}
\noindent
The first two terms constitute an inductive voltage divider: $\mathbf{v}_2$ is a weighted average of the capacitor voltage $\mathbf{v}_c$ and the grid voltage $\mathbf{v}_g$, with weights $L_g/L_t$ and $L_2/L_t$ respectively. The third term is a small resistive correction that vanishes when the X/R ratios of the two segments are matched.

\textit{Remark:}
The divider ratio $\gamma \triangleq L_2 / L_t$ governs how much of the grid-source disturbance propagates to the POI. On a weak grid ($L_g \gg L_2$), $\gamma \to 0$ and the POI is effectively pinned to $\mathbf{v}_c$; on a stiff grid ($L_g \to 0$), $\gamma \to 1$ and the measurement collapses onto $\mathbf{v}_g$.

\subsection{Phase Jump as an Initial-Condition Perturbation}
\label{sec:phase_jump}

The phase-angle jump is modeled as an instantaneous step $\Delta\theta$ in the infinite-bus voltage angle. The synchronous reference frame, aligned with the infinite bus voltage, goes through the same angle step. The physical states are continuous through the jump, i.e., inductor currents and capacitor voltages do not change instantaneously, but the coordinate transformation from the pre-jump to the post-jump frame introduces a passive rotation. Defining
\begin{equation}
  \mathbf{R}(\Delta\theta) = \begin{bmatrix}
    \cos\Delta\theta & \sin\Delta\theta \\
    -\sin\Delta\theta & \cos\Delta\theta
  \end{bmatrix},
  \label{eq:rot}
\end{equation}
a phase jump of $\Delta \theta$ at $t=0$ is modeled by a mere initial condition in the post-jump frame as 
\begin{equation}
  \mathbf{x}(0^+) = \mathrm{diag}\bigl(\mathbf{R},\,\mathbf{R},\,\mathbf{R}\bigr)\;\mathbf{x}_{\mathrm{ss}},
  \label{eq:x0}
\end{equation}
\noindent
with $\mathbf{R} = \mathbf{R}(\Delta\theta)$ applied to each $(d,q)$ pair, where $\mathbf{x_{ss}}$ denotes the pre-disturbance steady-state. Thus, the entire disturbance enters the problem as a deterministic rotation of the equilibrium state; no exogenous signal acts for $t > 0$.

\subsection{Instantaneous Power at Three Measurement Points}
\label{sec:power_meas}
Active power can be evaluated at three nodes of interest---$P_c$ at the inverter terminal, $P_{poi}$ at the POI, and $P_g$ at the infinite bus from the inner product of the respective voltage and current vectors ($P=\mathbf{v}\cdot\mathbf{i}$).

\textit{Proposition 1 (Power-drops at $t=0^+$):}
Immediately after a phase jump $\Delta\theta$, and before any control action takes effect:

\noindent (i)\; $P_{\mathrm{c}}(0^+) = P_{\mathrm{c}}(0^-)$ \textup{(invariant)};

\noindent (ii)\; $P_g(0^+) = \cos\Delta\theta \cdot P_{g}(0^-)$;

\noindent (iii)\; $P_{\mathrm{poi}}(0^+) - P_{\mathrm{poi}}(0^-) = -\gamma\cdot(1 - \cos\Delta\theta)\cdot P_{\mathrm{g}}(0^-)$.

\smallskip
\textit{Proof:}
Statement (i) follows from the orthogonality of $\mathbf{R}$: both $\mathbf{v}_c$ and $\mathbf{i}_2$ are rotated by the same matrix, and the inner product $\mathbf{v}_c^\top \mathbf{i}_2 = \mathbf{v}_c^\top \mathbf{R}^\top \mathbf{R}\,\mathbf{i}_2$ is preserved. Following the phase jump, $v_{gd}(0^+)=v_{gd}(0^-)$, $v_{gq}(0^+)=v_{gq}(0^-)=0$, $i_{2d}(0^+)=\cos \Delta \theta \cdot i_{2d}(0^-)$, and hence Statement (ii) immediately follows. Using \eqref{eq:vmeas}, power at POI is obtained as:
\begin{equation*}
  \begin{aligned}
    P_{\mathrm{poi}}(0^-) &= \tfrac{L_g}{L_t}\underbrace{\mathbf{v}_c(0^-)^\top \mathbf{i}_2(0^-)}_{= P_{\mathrm{c}}(0^-)} + 
    \tfrac{L_2}{L_t}\underbrace{\mathbf{v}_g^\top \mathbf{i}_2(0^-)}_{= P_{\mathrm{g}}(0^-)} + \epsilon\\
    P_{\mathrm{poi}}(0^+) &= \tfrac{L_g}{L_t}\underbrace{\mathbf{v}_c(0^+)^\top \mathbf{R}^\top \mathbf{R}\mathbf{i}_2(0^-)}_{= P_{\mathrm{c}}(0^-)} + 
    \tfrac{L_2}{L_t}\underbrace{\mathbf{v}_g^\top \mathbf{R}\mathbf{i}_2(0^-)}_{= \cos\Delta\theta \cdot P_{\mathrm{g}}(0^-)} + \epsilon    
  \end{aligned}
\end{equation*}
where $\epsilon=(L_2 R_g - L_g R_2)/L_t \cdot |\mathbf{i}_2|^2$ denoting the resistive correction term, and statement (iii) immediately follows.
\hfill$\blacksquare$

Proposition~1 reveals that the power remains invariant at the inverter terminal, however the infinite bus experiences an 
inevitable drop in active power immediately following the phase jump, whereas the POI experiences a smaller but similar 
drop defined by the ratio $\gamma = L_2/(L_2+L_g)$. For instance, an $L_2=0.15$pu (MVT reactance typically ranges 
between $5\%-7\%$, whereas GSUs exhibit $8\%-15\%$) yields $\gamma \approx 0.31$ 
under the ERCOT AGS-ESR test scenario (SCR\,=\,3, $X/R$\,=\,6); a $-25^\circ$ jump produces a 9.4\% drop at the 
infinite bus and a 2.9\% drop at the measurement point, i.e., POI. This has a direct implication for 
interpreting compliance test results: any observed instantaneous power drop, specially for negative phase jumps, 
should not be interpreted as disqualifying since such an excursion is a consequence of physics of the 
disturbance itself and not attributable to any control action by the inverter. Note that compliance tests typically allow 
low-pass filters on the measurement with small time-constants ranging from one to several ms, which can obscure such instantaneous power drops.

\section{Ideal Grid-Forming Response}
\label{sec:ideal_gfm}

\subsection{Grid-Side Current Dynamics of an Ideal Voltage Source}
\label{sec:grid_dynamics}
On the time scale of the phase-jump transient (the first 10--30\,ms after $t = 0^+$), an ideal grid-forming inverter is expected to behave as an ideal AC voltage source holding its terminal voltage at the post-jump initial condition $\mathbf{v}_c(t) = \mathbf{v}_c(0^+) = \mathbf{R}(\Delta\theta)\, \mathbf{v}_{c,ss}$, since the droop or virtual swing-equation dynamics, operating on the electromechanical time scale, have barely begun to act. Under this stiff voltage-source condition, the grid-side current $\mathbf{i}_2 = [i_{2d}\;\; i_{2q}]^\top$ is driven through the lumped impedance $(L_t, R_t)$ by the constant voltage mismatch $\mathbf{v}_c(0^+) - \mathbf{v}_g$. From \eqref{eq:di2}:
\begin{equation}
\begin{split}
    \dot{\mathbf{i}}_2 =& A_g\,\mathbf{i}_2 + \mathbf{b}_g,\\
    A_g = \begin{bmatrix}
    -\sigma & \omega_0 \\[2pt]
    -\omega_0 & -\sigma
  \end{bmatrix}\!,\quad
  &\mathbf{b}_g = \frac{\omega_b}{L_t}
  \begin{bmatrix}
    v_{cd}(0^+) - v_{gd} \\[2pt]
    v_{cq}(0^+) - v_{gq}
  \end{bmatrix}
\end{split}
  \label{eq:i2_frozen}
\end{equation}
\noindent
with $\sigma \triangleq \omega_b R_t / L_t$ denoting the electromagnetic damping rate. The eigenvalues $\lambda_{1,2} = -\sigma \pm j\omega_0$ yield the closed-form solution:
\begin{equation}
  \mathbf{i}_2(t) = \mathbf{i}_{2,\infty}
    + e^{-\sigma t}\,\mathbf{R}(-\omega_0 t)\,
      \bigl(\mathbf{i}_2(0^+) - \mathbf{i}_{2,\infty}\bigr),
  \label{eq:i2_soln}
\end{equation}
where $\mathbf{i}_{2,\infty} = -A_g^{-1}\mathbf{b}_g$ is the equilibrium under frozen voltage and $\mathbf{R}(\cdot)$ is the rotation matrix defined in~\eqref{eq:rot}. For a negative phase jump ($\Delta\theta < 0$), the angle $\delta = \angle\mathbf{v}_c - \angle\mathbf{v}_g$ increases, so $i_{2d,\infty} > i_{2d}(0^-)$: the frozen voltage source pushes more power into the grid.

The damped $\omega_0$ oscillation in~\eqref{eq:i2_soln} is the dq-frame manifestation of the classical DC offset in a switched RL circuit; its frequency and damping are set entirely by $(L_t, R_t)$, not by the inverter's filter design, and no controller can suppress it without deviating from ideal voltage-source behavior by modulating $\mathbf{v}_c$~\cite{rokrok2022transient, qoria2020current}. Consequently, expecting a monotonic active-power response to a phase jump implicitly requires the inverter to sacrifice stiff voltage-source fidelity in favor of electromagnetic damping---a tradeoff that the OCP in Section~\ref{sec:ocp} quantifies.

\subsection{Peak Current and Minimum Overload for Ideal Behavior}
\label{sec:overload}
\begin{figure}[htb]
	\makebox[\linewidth][c]{\includegraphics[angle = 0, clip, trim=0cm 0cm 0cm 0cm, width=0.5\textwidth]{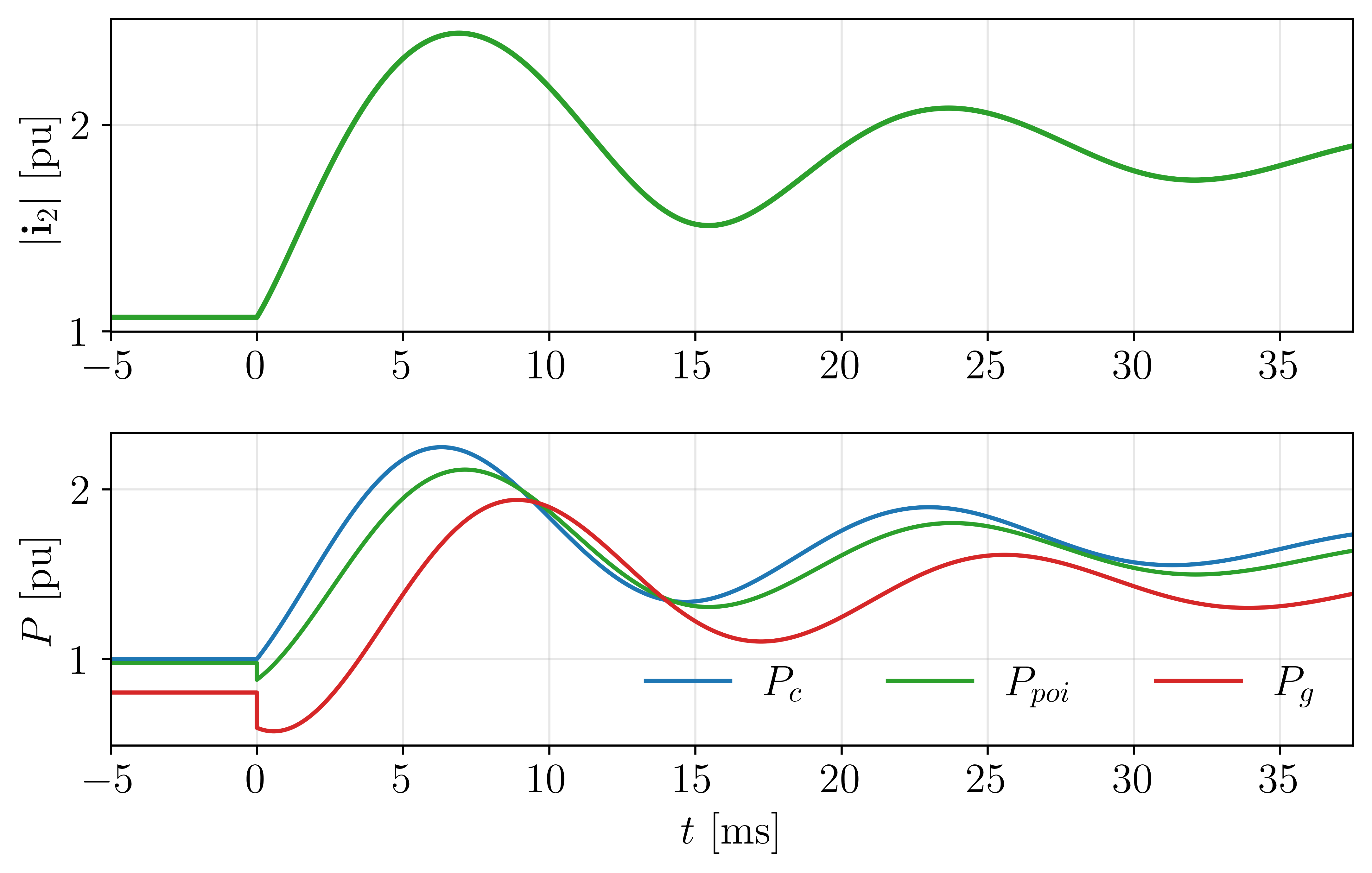}}
	\caption{Ideal GFM response to a $-25^\circ$ phase jump without capacity constraints.}
	\label{fig:ideal_gfm}
\end{figure}
The peak output current over the entire transient defines the
minimum overload threshold for ideal voltage-source behavior:
\begin{equation}
  I_{\mathrm{max}}^{\mathrm{ideal}}(\Delta\theta) \;=\;
  \max_{t \ge 0}\;\bigl|\mathbf{i}_2(t)\bigr|
  \label{eq:Imax_ideal}
\end{equation}
where $\mathbf{i}_2(t)$ is the closed-form solution 
\eqref{eq:i2_soln}. The maximum is attained near 
$t \approx \pi/\omega_0$ (half a line cycle), when the decaying 
spiral reaches its outermost excursion. 
Equation~\eqref{eq:Imax_ideal} is fully analytical and 
control-architecture independent: it depends only on the 
eigenvalues of $A_g$, the grid impedance, and the pre-disturbance 
steady state.

Fig.~\ref{fig:ideal_gfm} depicts a representative ideal GFM response 
to a $-25^\circ$ jump in absence of any capacity constraints for SCR of 3 and X/R of 6. The grid-side current and all
three power signals exhibit the damped $\omega_0$ oscillation
derived in Section~\ref{sec:grid_dynamics}.  The active power at the
inverter terminal $P_c$ \emph{rise} immediately after the jump: 
the increased angle $\delta = \angle\mathbf{v}_c - \angle\mathbf{v}_g$ 
drives additional power through the grid impedance.  The power continues to rise
monotonically during the first few milliseconds, then overshoots the respective
new equilibrium and oscillates around it at line frequency. 
The grid-source power $P_g = i_{2d}$ follows the same
pattern but with a brief initial dip below $\cos\Delta\theta\cdot
P_{\mathrm{pre}}$ (as identified in Proposition~1) before joining
the rising trajectory.  At the POI, this initial dip is attenuated
by the voltage-divider ratio $\gamma$ and may largely be obscured by
the measurement filter (with time-constants up to a ms).

\section{Implications for Current-Limited Operation}
\label{sec:transition}
Regulators rightly recognize that the overload levels identified 
above are impractical and provide carve-outs for current-limited 
operation~\cite{ercot_ags_faq}. The ERCOT AGS-ESR specification, 
for instance, acknowledges that when the inverter reaches its 
current limit, the standard active-power criteria ``may not be 
applicable,'' but still expects that ``power must return to the 
pre-disturbance level in a stable manner''~\cite{ercot_ags_faq}---
interpreted in practice as a monotonic rise in $P_{\mathrm{poi}}$ 
followed by a controlled decay. This relaxed expectation still embeds a nontrivial hardware 
requirement. Under ideal GFM behavior, $P_{\mathrm{poi}}$ rises 
monotonically while $|\mathbf{i}_2| < I_{\max}$, where $I_{\max}$ denotes the allowable current limit by the inverter. The moment $|\mathbf{i}_2|$ reaches $I_{\max}$, the controller must deviate from strict 
voltage regulation to enforce the constraint. If the inverter lacks 
sufficient headroom, this limiting action engages while 
$P_{\mathrm{poi}}$ is still close to its pre-disturbance value, and 
the resulting voltage deviation can cause $P_{\mathrm{poi}}$ to dip 
below $P_{\mathrm{poi}}(0^-)$ before recovering---precisely the 
behavior the specification seeks to avoid. Whether a dip occurs 
depends on the available current margin (a hardware parameter), the 
abruptness of the limiting transition, and the partitioning of the 
limited current between active and reactive 
axes~\cite{rokrok2022transient} (both controller design choices).

This motivates the central question: given a fixed $I_{\max}$, what 
is the best achievable $P_{\mathrm{poi}}$ trajectory across all 
possible GFM architectures and limiter strategies? If even the 
optimal trajectory---solved offline with perfect disturbance 
knowledge---exhibits a dip below $P_{\mathrm{poi}}(0^-)$, then no 
realizable controller can avoid it, and the specification implicitly 
requires a higher overload ratio. The minimum $I_{\max}$ at which 
the dip vanishes, denoted $I_{\max}^{\mathrm{mono}}$, is the 
practical overload mandate embedded in the grid-code language and is 
the central quantity computed in the following section.
In the analysis that follows, the overload capability is formulated in terms of the terminal 
current $|\mathbf{i}_2|$ rather than the inverter-side current 
$|\mathbf{i}_1|$. While the physical semiconductor limit applies 
to $|\mathbf{i}_1|$, inverter-side filter topologies vary across 
vendors, and the mapping from a given $|\mathbf{i}_1|$ limit to the 
corresponding $|\mathbf{i}_2|$ boundary is vendor-specific. 
Expressing the constraint at the terminal abstracts the filter 
design and makes the analysis universally applicable: for any 
specific hardware, the manufacturer's declared terminal current 
capability $I_{\max}$ can be used directly.

\section{Constrained Optimal Control with GFM Voltage-Tracking Objective}
\label{sec:ocp}
Any causal grid-forming controller is expected to produce an identical output, the frozen-voltage-natural-response given by \eqref{eq:i2_soln}, until the output current reaches a threshold level $I_{\mathrm{th}}$, where $I_{\mathrm{th}}$ denotes the operating limit above which a GFM controller engages its output limiting action. This threshold is set above continuous rating with a reasonable margin to avoid spurious activation during normal load transients, e.g., $I_{\mathrm{th}}=1.2$pu. Beyond the instant $t=t_{\lim}$, when $|\mathbf{i}_2|$ reaches $I_{\mathrm{th}}$, the controller must deviate from ideal voltage-source behavior to enforce the current constraint. This section formulates the \emph{best achievable} voltage-tracking trajectory from $t_{\lim}$ onward as a constrained optimal control problem (OCP), producing a controller-architecture-independent physical bound on the achievable power-recovery performance.

\subsection{Two-Phase Trajectory Structure}
The post-jump trajectory decomposes into two phases. In \textbf{Phase~1} ($0 \leq t < t_{\lim}$), below the current limit the inverter approximates an ideal voltage source keeping $\mathbf{v}_c(t) = \mathbf{v}_c(0^+)$; the grid-side current follows~\eqref{eq:i2_soln}. Every causal GFM controller, regardless of its architecture or current-limiting strategy, should produce the same trajectory because no measurable change has yet prompted a control action. The transition instant
\begin{equation}
    t_{\lim} \triangleq \inf\left\{t > 0 : 
    \left|\mathbf{i}_2(t)\right| = I_{\mathrm{th}}\right\},
    \label{eq:tlim}
\end{equation}
marks when the current first reaches the limiter engagement threshold $I_{\mathrm{th}}$, set above continuous rating to avoid spurious activation (e.g., $I_{\mathrm{th}} = 1.2$\,pu). In \textbf{Phase~2} $t \geq t_{\lim}$), the controller must modify $\mathbf{v}_c$ to enforce $|\mathbf{i}_2| \leq I_{\max}$. The optimal trajectory from this point onward is the solution to the OCP formulated below. The initial condition for \textbf{Phase 2} is fully determined by~\eqref{eq:i2_soln} evaluated at $t_{\lim}$.

\subsection{Grid-Forming Objective}
\label{sec:gfm_objective}
A GFM controller regulates terminal voltage, not power. On the phase-jump time scale, the droop or VSM dynamics adjust the voltage reference from its frozen value toward the post-disturbance equilibrium. As a representative approximation across various GFM control structures, this re-synchronization can be characterized by a time constant $\tau_{\text{sync}}$ (typically 20--100\,ms). The resulting voltage reference trajectory in the post-jump frame is modeled as a first-order relaxation:
\begin{equation}
    \mathbf{v}_{c,\text{ref}}(t) = \mathbf{v}_{c,ss} 
    + \left(\mathbf{v}_{c}(0^+) - \mathbf{v}_{c,ss}\right) 
    e^{-t/\tau_{\text{sync}}},
    \label{eq:vc_ref}
\end{equation}
where $\mathbf{v}_{c,ss}$ is the post-disturbance steady state 
(identical to the pre-disturbance value since the grid-voltage 
magnitude is unchanged) and 
$\mathbf{v}_c(0^+) = \mathbf{R}(\Delta\theta)\,\mathbf{v}_{c,ss}$. 
The voltage-tracking objective is:
\begin{equation}
    J = \int_{0}^{T} 
    \left\|\mathbf{v}_c(t) - \mathbf{v}_{c,\text{ref}}(t)\right\|^2 
    dt,
    \label{eq:obj}
\end{equation}
which penalizes deviations in both magnitude and angle simultaneously. Alternative weightings may yield different trajectories, but a GFM controller cannot distinguish event types 
in real time and must use a single control law across phase jumps, voltage sags, load steps, and faults alike. Any magnitude-versus-frequency prioritization must therefore be justified across the full spectrum of ride-through scenarios; no consensus or empirical evidence exists in favor of any such prioritization. Equal weighting provides a neutral, architecture-agnostic baseline.

\subsection{Optimal Control Formulation}
To generalize the optimal control formulation, we isolate the reduced dynamics defined in Section~\ref{sec:grid_dynamics}, treating the internal capacitor voltage $\mathbf{v}_c$ as the effective control input to the grid-side impedance. This analytical formulation implicitly assumes a near-infinite control bandwidth over the inverter output voltage, establishing the resulting optimal trajectory 
as the absolute maximum performance bound achievable by any voltage control structure.

The Phase~2 OCP is stated as follows.
\begin{equation}
\begin{aligned}
    \min_{\mathbf{v}_c(\cdot)} \quad 
    & \int_{0}^{T} \left\|\mathbf{v}_c - \mathbf{v}_{c,\text{ref}}
    \right\|^2 dt + w_T\,\Phi\left(\mathbf{i}_2(T)\right) 
    + w_r \int_0^T \|\dot{\mathbf{v}}_c\|^2 dt \\
    \text{s.t.} \quad 
    & \dot{\mathbf{i}}_2 = A_g\,\mathbf{i}_2 + B_g\,\mathbf{v}_c 
      + \mathbf{d}, \quad
      \mathbf{i}_2(0) = \mathbf{i}_2(t_{\lim}), \\
    & |\mathbf{i}_2(t)|^2 \leq I_{\max}^2, \quad 
      |\mathbf{v}_c(t)|^2 \leq V_{\max}^2, \quad
      \forall\, t \in [0, T],
\end{aligned}
\label{eq:ocp}
\end{equation}
where $\Phi(\mathbf{i}_2(T)) = (\mathbf{i}_2(T) - 
\mathbf{i}_{2,ss})^\top Q_T(\mathbf{i}_2(T) - \mathbf{i}_{2,ss})$, $Q_T \succ 0$ defining a terminal penalty with weight $w_T$ driving the state toward the pre-disturbance operating point, and $w_r$ weights a control-rate regularizer that suppresses unrealizable 
high-frequency chattering.

\textit{Remark (Scope of the OCP):}
The OCP treats the terminal voltage $\mathbf{v}_c$ as a free 
control input, implicitly assuming infinite inner-loop bandwidth 
and bypassing converter-side constraints such as the inverter 
current $|\mathbf{i}_1|$, DC bus voltage, and modulation limits. 
The resulting $I_{\max}^{\text{mono}}$ is therefore a 
\emph{lower bound} on the true overload requirement: any 
real hardware, subject to additional topology-dependent 
constraints (power-stage architecture, DC-side 
dynamics) would require equal or greater overload capability. 
This generality is deliberate---it ensures the bound applies 
across vendor-specific designs and power-stage topologies without loss of validity.
A vendor-specific formulation with $\mathbf{v}_i$ as the 
control input and explicit constraints on $|\mathbf{i}_1|$ 
and $V_{\text{dc}}$ would yield tighter but hardware-dependent 
bounds. Similarly, the frozen-voltage assumption in Phase~1 
represents ideal causal GFM behavior: on sufficiently fast time-scales compared to that of the primary frequency response, a well-designed 
controller operating within its current capability has no 
reason to deviate from its terminal voltage reference, as 
doing so would require anticipating a disturbance not yet 
reflected in any measurable quantity.

\subsection{Convexity and Global Optimality}
The OCP~\eqref{eq:ocp} is a convex optimization problem: the 
plant dynamics~\eqref{eq:i2_frozen} are 
linear in the state $\mathbf{i}_2$ and control $\mathbf{v}_c$, 
the current and voltage constraints define convex (Euclidean ball) 
feasible sets in the state and control spaces respectively, and 
both the running cost and terminal penalty are convex quadratics. 
Convexity is preserved under time discretization, yielding a 
quadratically constrained quadratic program (QCQP) in which any 
local minimum is necessarily the global minimum. This property is 
essential to the paper's central argument: because the offline 
solution is globally optimal, it constitutes an absolute bound 
that no online controller can surpass.

\subsection{Numerical Solution} \label{sec:ocp_numerical_soln}
While the OCP is mathematically convex, direct transcription in Cartesian $(i_{2d}, i_{2q})$ coordinates leads to poor numerical conditioning due to the $\pm\omega_0$ cross-coupling and the quadratic current constraint being active over extended intervals. A polar transformation $I_2 = |\mathbf{i}_2|$, $\varphi = \angle\mathbf{i}_2$ reduces the current constraint to the elementary variable bound $I_2 \leq I_{\max}$, with dynamics:
\begin{align}
    \dot{I}_2 = \cos\varphi\,\dot{i}_{2d} 
    + \sin\varphi\,\dot{i}_{2q},\ 
    \dot{\varphi} = \frac{\cos\varphi\,\dot{i}_{2q} 
    - \sin\varphi\,\dot{i}_{2d}}{I_2}
\end{align}
where $\dot{i}_{2d}$ and $\dot{i}_{2q}$ follow from~\eqref{eq:i2_frozen}. The singularity at $I_2 = 0$ is excluded by a small lower bound, never active since Phase~2 begins at $I_2(0) = I_{\mathrm{th}}$.

The OCP~\eqref{eq:ocp} in polar coordinates is transcribed into a finite-dimensional nonlinear program (NLP). The horizon $[0, T]$ is divided into $N$ intervals, with dynamics propagated via fourth-order Runge--Kutta, yielding $2(N{+}1)$ state variables, $2N$ control variables, $N{+}1$ variable bounds, and $N$ quadratic constraints. The NLP is solved using IPOPT~\cite{wachter2006ipopt} via CasADi~\cite{andersson2019casadi}. With discretization rates substantially exceeding the sampling and switching frequencies of commercial inverters, the solution represents the true physical bound: no realizable GFM controller can track the voltage reference 
more closely while respecting the same hardware constraints.

In the following analysis, the various weights in \eqref{eq:ocp} are set as $w_T = 10$, $Q_T = \text{diag}(q_{|\mathbf{i}|},\;q_{\angle{\mathbf{i}}})$ (in polar coordinates, penalizing current-magnitude error twice as heavily as angle error), and $w_r = 10^{-4}$; sensitivity analysis, such as $w_T=\{5,10,15\}$, $q_{|\mathbf{i}|}=[0.5, 1]$, $q_{\angle{\mathbf{i}}}=[0.5, 1]$, confirmed that the computed $I_{\max}^{\text{mono}}$ is insensitive to moderate variations in these weights, as the result is dominated by the hard current constraint rather than the soft penalties. The re-synchronization time-constant is set as $\tau_{\textbf{sync}}=20$ms and the optimization horizon is set to $T = 3\tau_{\textbf{sync}}$, ensuring that secondary oscillatory swings in the power trajectory are captured. Increasing $T$ to $6\tau_{\textbf{sync}}$ yields $I_{\max}^{\text{mono}}$ values within 2.3\% for all phase-jump magnitudes tested.

\begin{figure}[tb]
	\makebox[\linewidth][c]{\includegraphics[angle = 0, clip, trim=0cm 0cm 0cm 0cm, width=0.35\textwidth]{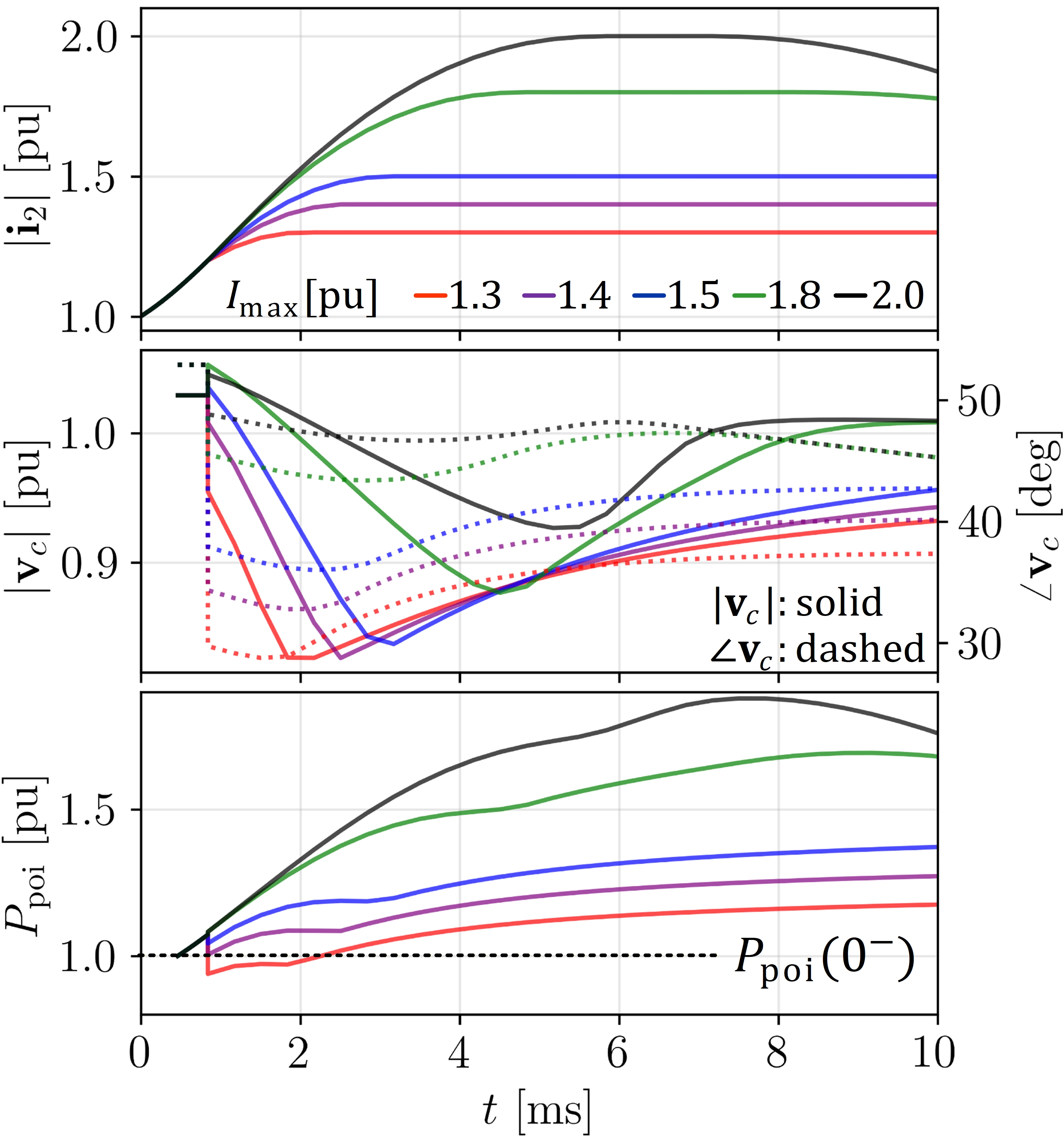}}
	\caption{Optimal output trajectory in Phase 2 for different overload limits in response to a $-25^\circ$ phase jump; $I_{\text{th}}$ kept at 1.2 pu.}
	\label{fig:ocp_imax_sweep}
\end{figure}

Fig.~\ref{fig:ocp_imax_sweep} depicts the optimal trajectories in response to a $-25^\circ$ phase jump, assuming a threshold current of $I_{\mathrm{th}} = 1.2$ pu. As established in Section~\ref{sec:overload}, the power at the POI undergoes an inevitable momentary dip at $t=0^+$; the plotted trajectories begin only after the power recovers to its pre-disturbance level, $P_{\text{poi}}(0^-)$. Initially, for all evaluated $I_{\max}$ values, the trajectories perfectly track the ideal GFM response until the terminal current $|\mathbf{i}_2|$ reaches $I_{\mathrm{th}}$. Upon crossing this threshold, the controller must take immediate corrective action by adjusting the output voltage $\mathbf{v}_c$ (see $|\mathbf{v}_c|$ and $\angle{\mathbf{v}_c}$ trajectory in Fig.~\ref{fig:ocp_imax_sweep}) to strictly enforce the $I_{\max}$ hardware limit. Higher $I_{\max}$ ratings, such as 1.8 pu or 2.0 pu, provide a larger operating margin ($I_{\max} - I_{\mathrm{th}}$), resulting in a softer limiter engagement. Conversely, tighter limits force an abrupt pull-back in output power; for the strictest limit of 1.3 pu, the active power actually drops back below its pre-disturbance level to ensure hardware survival.

\subsection{Minimum Overload Ratio}
\label{sec:min_overload}
In this section, we define the \emph{minimum overload ratio} $I_{\max}^{\text{mono}}$ at which the Phase~2 power trajectory avoids any excursion below the pre-disturbance level $P_{\text{poi}}(0^-)$. This is computed by sweeping $I_{\max}$ and, for each value, solving the OCP~\eqref{eq:ocp} and recording
\begin{equation}
    P_{\min}^{\mathrm{ph2}}(I_{\max}) \triangleq \min_{t \geq t_{\lim}} P_{\text{poi}}(t),
    \label{eq:Pmin}
\end{equation}
\noindent
where $P_{\text{poi}}(t)$ is computed from the optimal trajectory via~\eqref{eq:vmeas}. The minimum overload is then obtained as
\begin{equation}
    I_{\max}^{\text{mono}} = \inf\left\{I_{\max} : P_{\min}^{\mathrm{ph2}}(I_{\max}) \geq P_{\text{poi}}(0^-)\right\}.
    \label{eq:Imono}
\end{equation}
\noindent
For $I_{\max} < I_{\max}^{\text{mono}}$, even the theoretically optimal GFM controller, with perfect foresight, unlimited computational power, and ideal voltage-source output, produces a transient power dip below $P_{\text{poi}}(0^-)$ during the current-limiting phase. Consequently, no realizable controller, regardless of its architecture or tuning, can avoid this drop. The power dip is a physical consequence of the current-limiting action itself, not a deficiency in control design. 

\begin{figure}[htb]
	\makebox[\linewidth][c]{\includegraphics[angle = 0, clip, trim=0cm 0cm 0cm 0cm, width=0.4\textwidth]{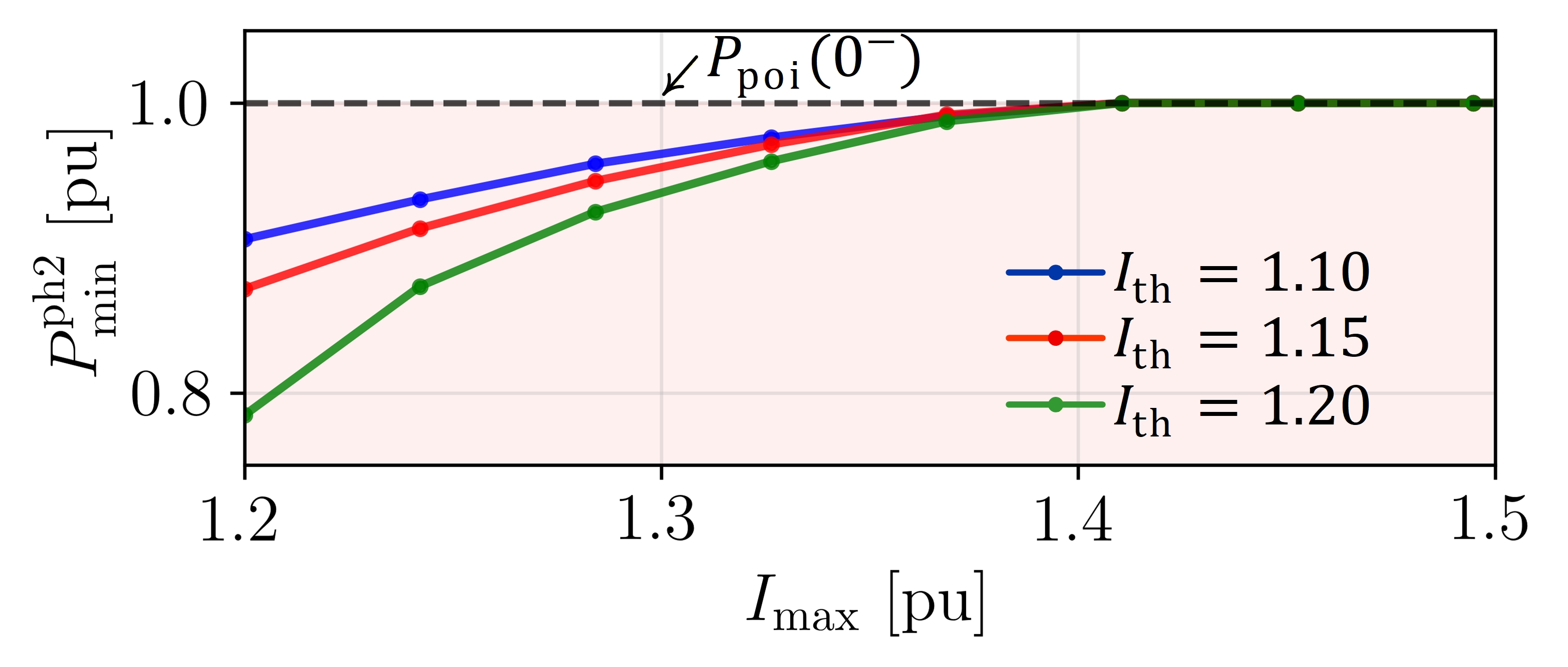}}
	\caption{Minimum active power output by optimal trajectories during Phase 2.}
	\label{fig:ocp_ppoi_min_p2}
\end{figure}

Fig.~\ref{fig:ocp_ppoi_min_p2} depicts the minimum power points traversed by the optimal trajectories for different $I_{\text{th}}$ values, which reveals a notable convergence: despite varying the engagement threshold, the critical overload $I_{\max}^{\text{mono}}$ at which the Phase~2 power dip vanishes is nearly identical across all three cases. This convergence is not a numerical artifact but reflects the geometric structure of the problem. At any instant during the transient, the POI power is governed by the projection of the 
current vector onto the grid-voltage axis, $P_{\text{poi}} \propto I_2 \cos\varphi$, 
and the minimum $I_{\max}$ required to maintain this projection above $P_{\text{poi}}(0^-)$ 
is dictated by the worst-case current angle $\varphi$ encountered along the 
trajectory, which is set by the phase-jump magnitude $\Delta\theta$ and the 
grid impedance $(L_t, R_t)$, not by the controller's engagement timing. A lower 
$I_{\text{th}}$ grants the optimal controller more lead time but from a lower 
initial power; a higher $I_{\text{th}}$ provides less lead time but a higher 
starting power. The optimal controller, possessing perfect foresight, exploits 
either scenario with equal effectiveness, so the two effects cancel. 
Consequently, $I_{\max}^{\text{mono}}$ is a \emph{plant parameter}, a function 
of $\Delta\theta$, SCR, $X/R$, and the pre-disturbance operating point, and 
is effectively independent of the detection threshold, reinforcing the paper's 
central contention that the overload mandate is embedded in the physics of the 
test configuration rather than in the controller design. It should be noted, 
however, that this threshold-independence is a property of the optimal bound; 
a causal controller without foresight would exhibit greater sensitivity to 
$I_{\text{th}}$, as it cannot compensate for a late engagement with anticipatory 
voltage adjustment.

\begin{figure}[htb]
	\makebox[\linewidth][c]{\includegraphics[angle = 0, clip, trim=0cm 0cm 0cm 0cm, width=0.4\textwidth]{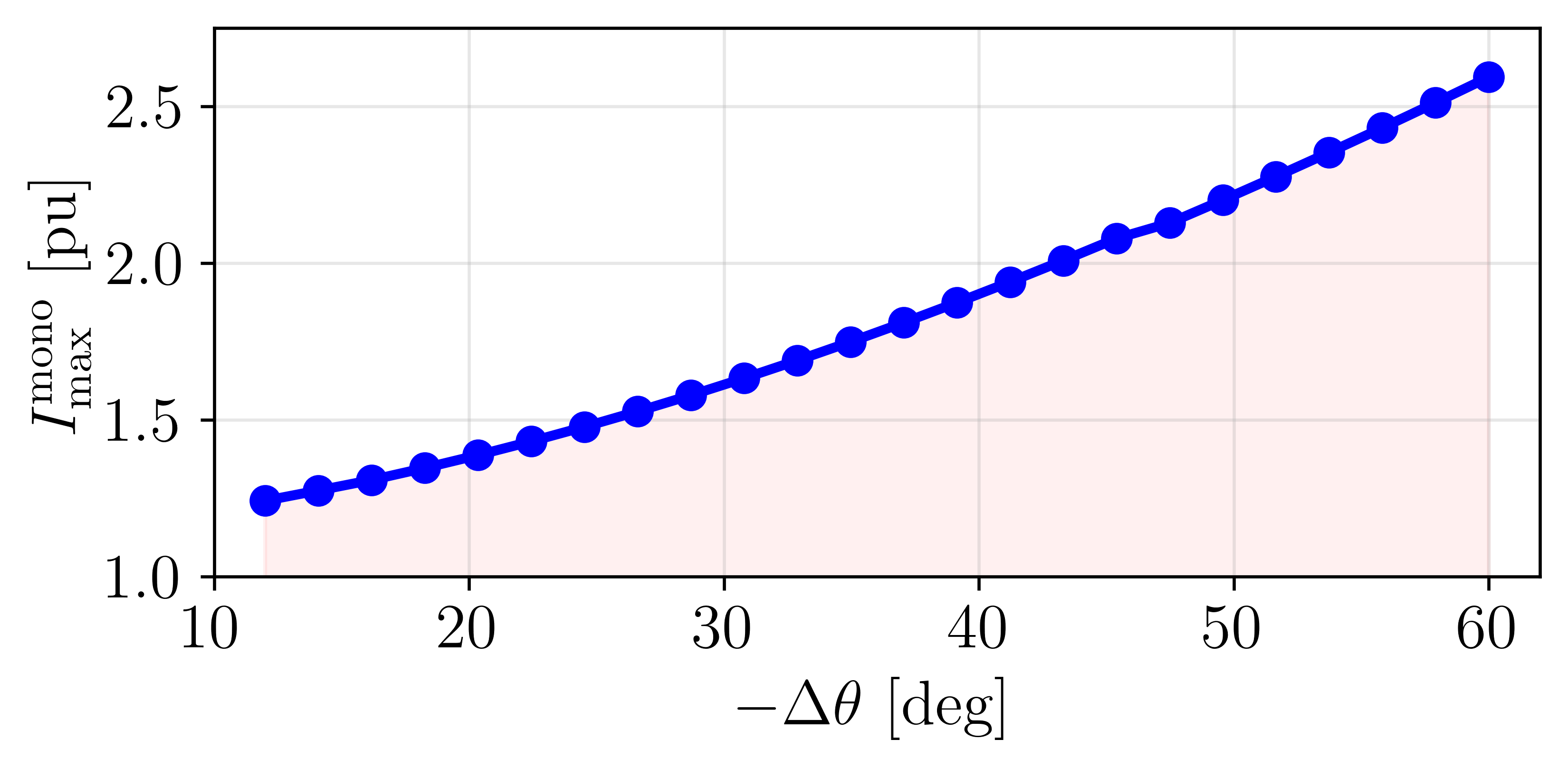}}
	\caption{Minimum overload required for monotonic recovery for various phase jump angles with $P_{\text{poi}}(0^-)=1$ pu.}
	\label{fig:imono_vs_angle}
\end{figure}

Fig.~\ref{fig:imono_vs_angle} extends this analysis by sweeping the 
phase-jump magnitude from $-10^\circ$ to $-60^\circ$ under the same 
test conditions ($P_{\text{poi}}(0^-) = 1$\,pu, SCR\,=\,3, $X/R$\,=\,6). 
The minimum overload $I_{\max}^{\text{mono}}$ rises monotonically 
with $|\Delta\theta|$, from approximately 1.24\,pu at $-10^\circ$ 
to 2.59\,pu at $-60^\circ$.

\section{EMT Simulation with Generic GFM Models}
The physical bound is validated in electromagnetic transient (EMT) simulation environment (PSCAD) against three WECC approved generic grid-forming inverter models~\cite{pnnl_regfm_pscad} developed under the UNIFI Consortium\cite{unifi_consortium}, each embodying a distinct GFM control philosophy and reflecting industry input from major inverter manufacturers:

\textbf{REGFM\_A1} (droop-based)~\cite{regfm_a1}: developed 
by PNNL with detailed input from SMA \cite{pnnl_regfm_pscad, du2024unifi}, based on the CERTS Microgrid 
Project~\cite{lasseter2010certs, du2018survivability}. The model implements 
$P$--$f$ and $Q$--$V$ droop control with anti-windup power limiters, and 
a virtual-impedance-based fault current limiter that switches from the 
droop-generated voltage reference $E_{\text{droop}} \angle \delta_{\text{droop}}$ 
to a current-constrained operating point when $|I| \geq I_{\max}$.

\textbf{REGFM\_B1} (virtual synchronous machine)~\cite{regfm_b1}: 
the initial model specification was proposed by PNNL, General Electric (GE), 
and the Electric Power Research Institute (EPRI), with additional input from 
Siemens Gamesa Renewable Energy (SGRE). The model emulates the swing equation 
dynamics of a synchronous generator, including configurable virtual inertia 
and damping, with a current magnitude limiter.

\textbf{REGFM\_C1} (hybrid GFL/GFM)~\cite{regfm_c1}: a 
GFL branch for steady-state dispatch paralleled with a 
VSM-based grid-forming branch for transient support; current 
limited by uniform algebraic scaling preserving the 
commanded phase angle. Developed by PNNL, Tesla Energy, and 
EPRI with WECC model validation subcommittee revisions.

The PSCAD model and respective documentations are available on github \cite{pnnl_regfm_pscad}. 
The diversity of contributors, spanning utility-scale inverter OEMs, national laboratories, and research institutes, ensures that the three models collectively represent the range of GFM control philosophies deployed or under development in the industry. This breadth strengthens the generality of the comparison against the 
physical bound.

\subsection{Simulation Setup}

The test follows Fig.~\ref{fig:topology} with SCR\,=\,3, $X/R$\,=\,6, 
$L_2 = 0.15$\,pu, $P_{\text{pre}} = 1.0$\,pu, $Q_{\text{pre}} = 0$. 
A phase jump $\Delta\theta = -25^\circ$ is applied at $t = 0$. Each 
model is simulated with $I_{\max}$ spanning 1.2--2.0\,pu, all other 
parameters kept at default values as found in \cite{pnnl_regfm_pscad}. $P_{\text{poi}}$ is measured per 
Section~\ref{sec:vmeas} and filtered with a 1\,ms smoothing constant 
per grid-code practice.

\subsection{Results and Discussion}
\label{sec:sim_results}

\begin{figure}[htb]
	\makebox[\linewidth][c]{\includegraphics[width=0.5\textwidth]{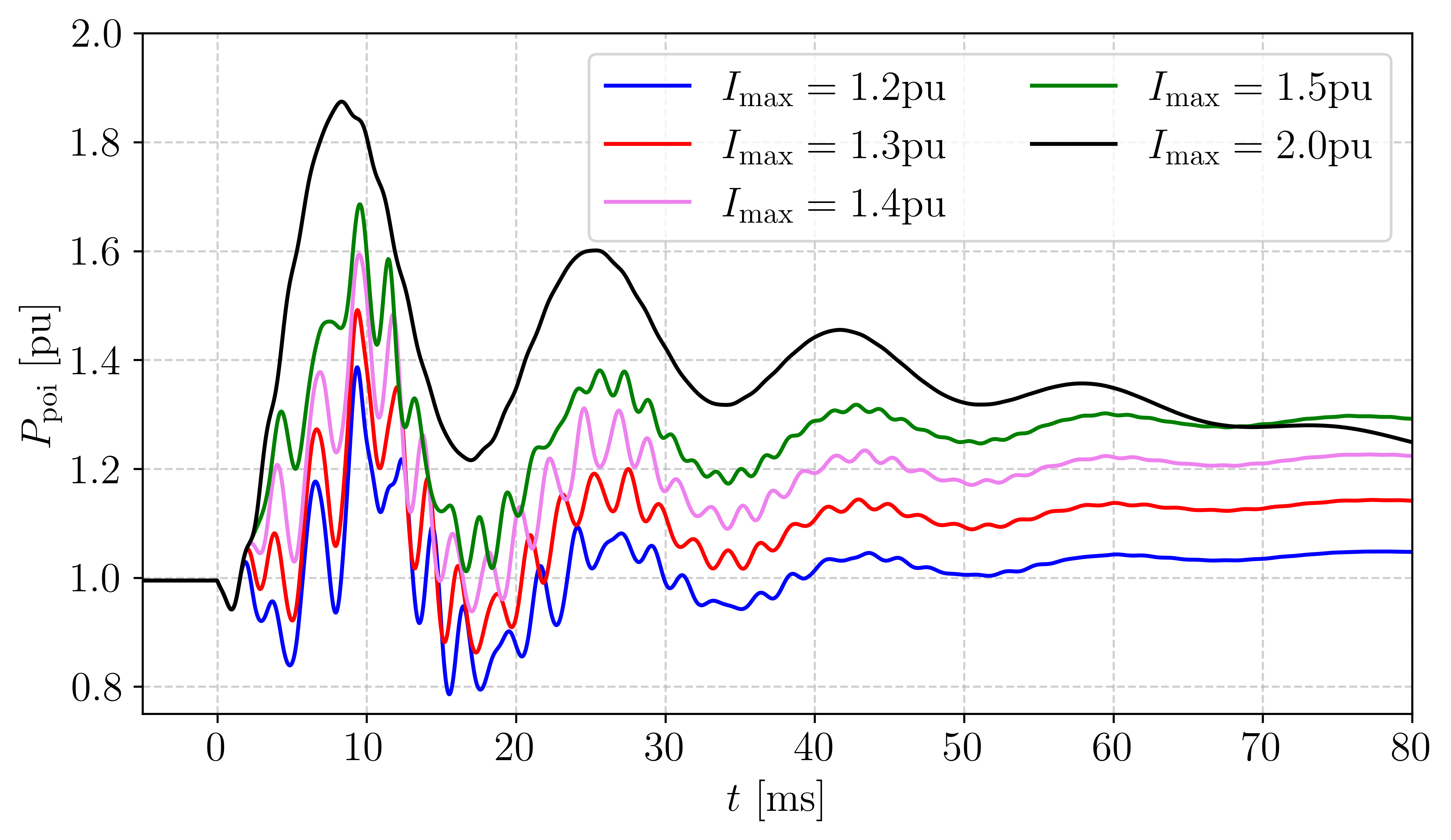}}
	\caption{$P_{\text{poi}}$ using REGFM\_A1 in response to $-25^\circ$ phase jump.}
	\label{fig:regfm_a1}
\end{figure}

\begin{figure}[htb]
	\makebox[\linewidth][c]{\includegraphics[width=0.5\textwidth]{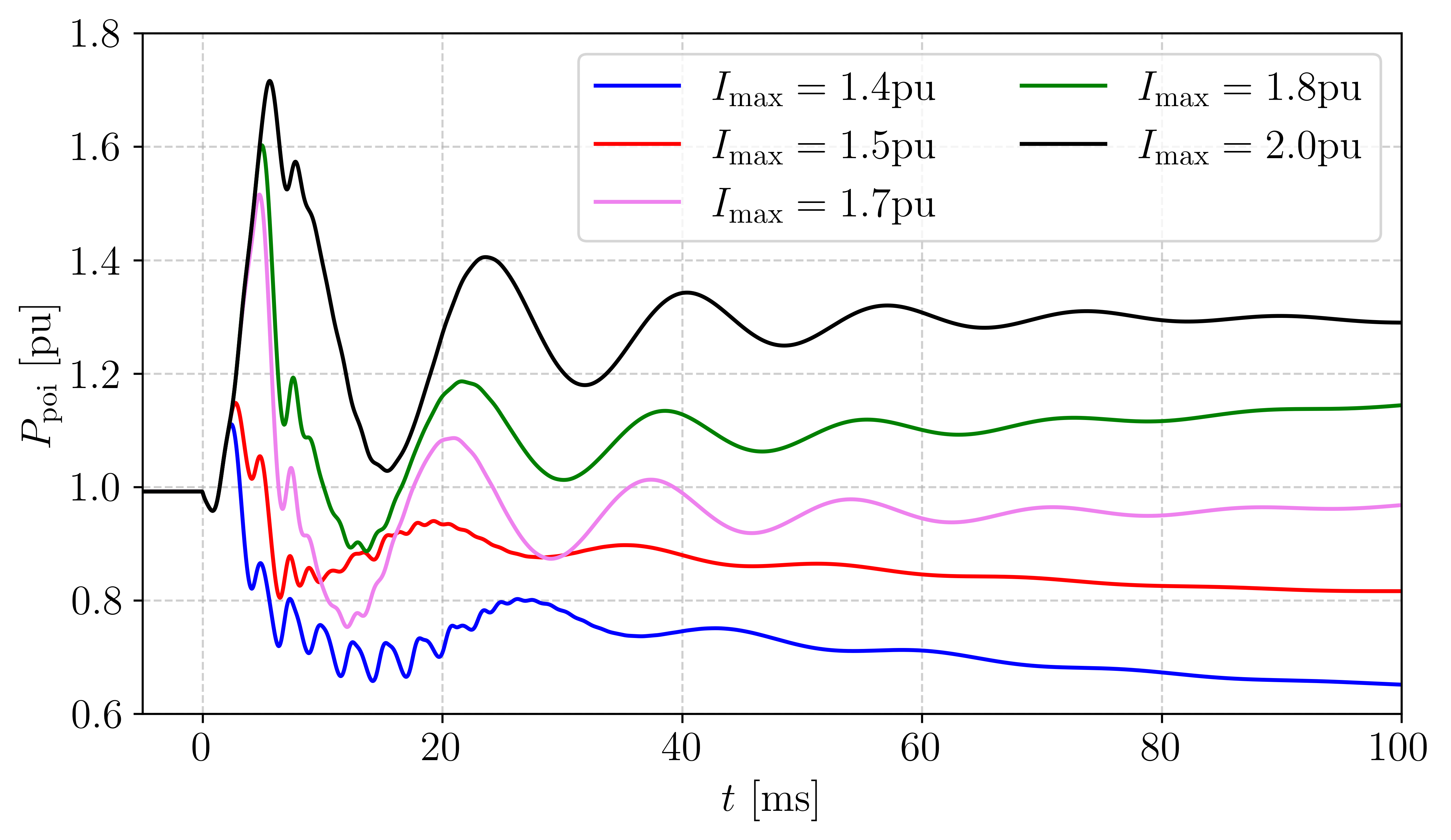}}
	\caption{$P_{\text{poi}}$ using REGFM\_B1 in response to $-25^\circ$ phase jump.}
	\label{fig:regfm_b1}
\end{figure}

\begin{figure}[htb]
	\makebox[\linewidth][c]{\includegraphics[width=0.5\textwidth]{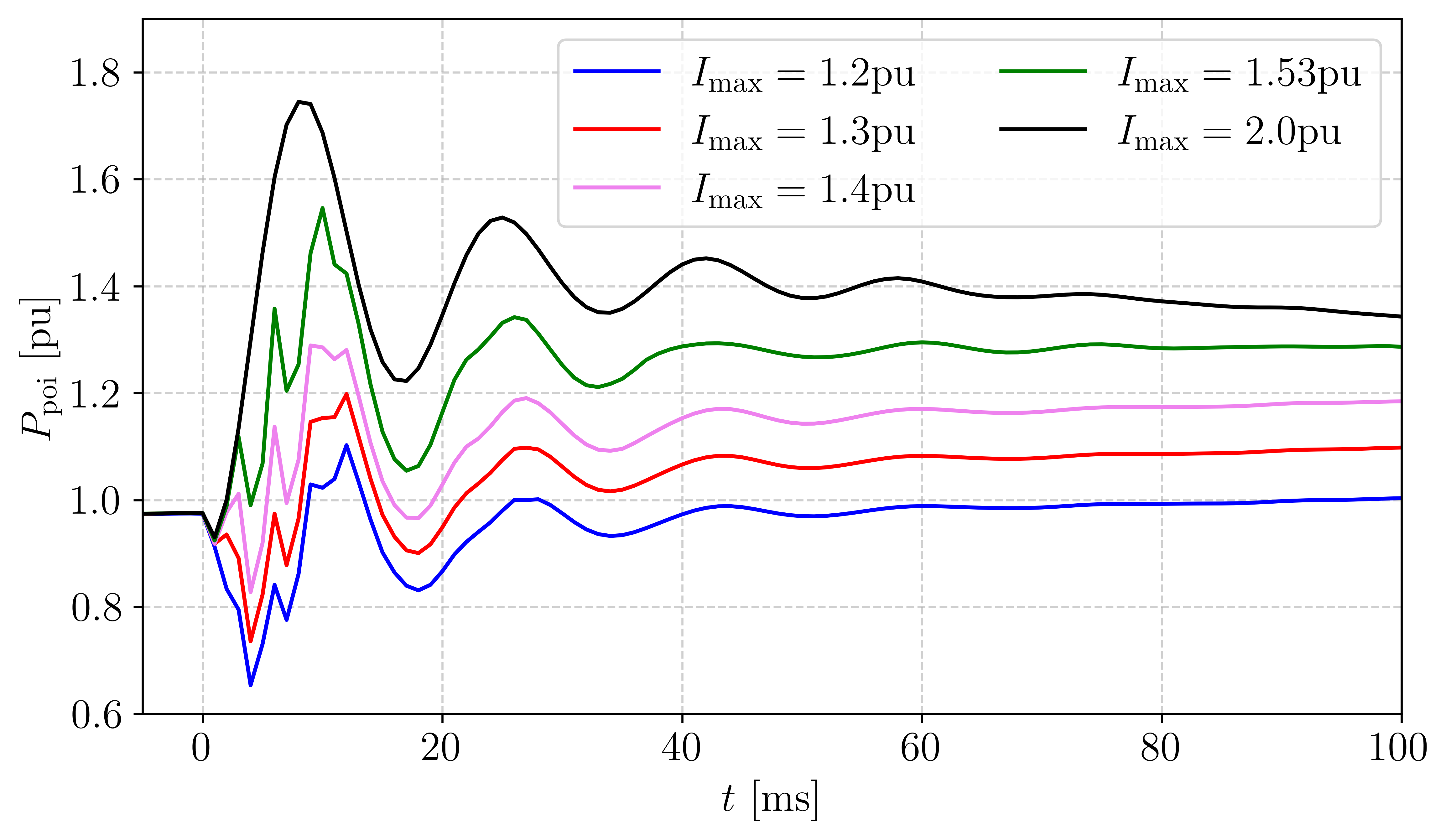}}
	\caption{$P_{\text{poi}}$ using REGFM\_C1 in response to $-25^\circ$ phase jump.}
	\label{fig:regfm_c1}
\end{figure}

Figs.~\ref{fig:regfm_a1}--\ref{fig:regfm_c1} present the POI power 
trajectories. All three models exhibit the momentary power drop at 
$t = 0^+$ predicted by Proposition~1.

\textbf{REGFM\_A1} (Fig.~\ref{fig:regfm_a1}) exhibits the 
line-frequency oscillation from Section~\ref{sec:ideal_gfm}, 
with higher-frequency content from the input filter that diminishes 
at higher $I_{\max}$. At $I_{\max} = 1.2$\,pu, $P_{\text{poi}}$ 
dips to $\approx$\,0.8\,pu; monotonic recovery above $P_{\text{pre}}$ 
is first achieved at $I_{\max} = 1.5$\,pu. At 2.0\,pu the response 
closely matches the unconstrained ideal trajectory of 
Fig.~\ref{fig:ideal_gfm}.

\textbf{REGFM\_B1} (Fig.~\ref{fig:regfm_b1}) shows the most severe 
limiter--synchronization interaction. At $I_{\max} = 1.4$\,pu, 
$P_{\text{poi}}$ drops to $\approx$\,0.65\,pu and fails to recover 
within 100\,ms, indicating potential loss of 
synchronism~\cite{rokrok2022transient}. A dip-free response requires 
$I_{\max} \geq 1.9$\,pu.

\textbf{REGFM\_C1} (Fig.~\ref{fig:regfm_c1}) achieves monotonic 
recovery at $I_{\max} \approx 1.53$\,pu with reduced oscillation but 
a sharper limiting transition. The contrast with B1 is instructive: 
both employ a VSM core, yet C1 requires substantially less overload. 
C1 limits current directly and algebraically (uniform scaling 
preserving phase angle), whereas B1 limits indirectly via 
integral-based regulation of the VSM angle saturation and voltage 
bounds. The limiting-loop lag in B1 allows the virtual rotor to 
accumulate excess energy before the current-limiting engages. This 
underscores that the \emph{speed of current enforcement}, not the 
synchronization philosophy, is the primary determinant of phase-jump 
recovery under saturation.

The $I_{\max}$ sweep ranges differ across the three models to 
bracket each model's respective $I_{\max}^{\text{mono}}$. For 
\textbf{REGFM\_B1}, values below 1.4\,pu are excluded as the model 
loses synchronization at these overload ratios, as evidenced by 
the non-recovery observed in Fig.~\ref{fig:regfm_b1}.

\subsection{Comparison with the Physical Bound}

\begin{table}[htb]
    \centering
    \caption{Minimum overload for monotonic $P_{\text{poi}}$ recovery 
    ($\Delta\theta = -25^\circ$, SCR\,=\,3, $X/R$\,=\,6)}
    \label{tab:imono_comparison}
    \begin{tabular}{lcc}
        \hline
        \textbf{Model / Bound} & $I_{\max}^{\text{mono}}$ (pu) & Gap \\
        \hline
        Physical bound (OCP) & 1.42 & --- \\
        REGFM\_A1 (droop) & $\approx$\,1.5 & +5.6\% \\
        REGFM\_C1 (hybrid) & $\approx$\,1.53 & +7.7\% \\
        REGFM\_B1 (VSM) & $\approx$\,1.9 & +33.8\% \\
        \hline
    \end{tabular}
\end{table}

Table~\ref{tab:imono_comparison} summarizes the results against the 
OCP bound from Section~\ref{sec:min_overload}. The bound is \emph{confirmed}: no architecture achieved monotonic recovered below 
$I_{\max}^{\text{mono}} = 1.42$\,pu. A realizable controller can \emph{closely approach} it: A1 and C1 
operate within 6--8\%, confirming that the voltage-tracking objective 
yields a practically attainable frontier. The gap is 
\emph{explainable}: models with direct current enforcement (A1, C1) 
approach the bound regardless of whether the synchronization is 
droop- or VSM-based; B1's indirect limiting incurs a 34\% penalty. 
The bound assumes stable recovery by construction and does not 
capture the loss-of-synchronism mode observed in B1 at low overload; 
it is a necessary but not sufficient condition for acceptable VSM 
performance.

\section{Implications for Grid-Code Development}

\subsection{Interpreting Compliance Test Results}

\subsubsection{The instantaneous power drop is physical, not a control failure}
Proposition~1 establishes that a momentary $P_{\text{poi}}$ drop 
immediately following a negative phase jump is an inevitable consequence 
of the coordinate rotation, governed by $\gamma = L_2/L_t$ and 
$\Delta\theta$. This drop is present even for an ideal, unconstrained 
voltage source (Fig.~\ref{fig:ideal_gfm}) and persists regardless 
of control architecture or current capability. Compliance evaluators 
should not interpret a sub-cycle excursion in the non-opposing direction 
as evidence of a control deficiency. Measurement filtering (time constants 
$\sim$1\,ms) may partially obscure but may not eliminate this excursion.

\subsubsection{Non-monotonic recovery may reflect current margin, not control quality}
When $P_{\text{pre}} \approx 1$\,pu and the phase jump drives the 
current to $I_{\max}$, the OCP analysis demonstrates that a transient 
dip below $P_{\text{pre}}$ is \emph{physically unavoidable} for 
$I_{\max} < I_{\max}^{\text{mono}}$, regardless of controller 
architecture. If the observed dip depth is comparable to the OCP 
prediction at the same $I_{\max}$ (within 10--15\%), the inverter is 
performing near the physical limit and no control improvement can 
meaningfully reduce the dip---the remedy is increased current capability. 
Conversely, a dip substantially deeper than the OCP prediction, or 
failure to recover within a reasonable window, indicates a structural 
limiter--synchronization interaction warranting investigation of the 
control design (cf.\ REGFM\_B1 in Section~\ref{sec:sim_results}).

\subsection{Specifying Test Conditions}

\subsubsection{Separate hardware limitations from control deficiencies}
A two-tier test structure is recommended:
\begin{enumerate}
    \item \textbf{Reduced-dispatch test} ($P_{\text{pre}} \leq 0.5$\,pu): 
    provides sufficient current margin for the full performance 
    criteria, such as peak change, rise time, sustained displacement, to 
    be evaluated without current-limit engagement, isolating 
    voltage-source fidelity from hardware limitations.
    \item \textbf{Rated-dispatch test} ($P_{\text{pre}} = 1.0$\,pu): 
    deliberately forces current limiting. Criteria should focus on 
    stable, damped recovery rather than monotonic rise, consistent 
    with existing relaxation provisions~\cite{ercot_ags_esr, ercot_ags_faq}.
\end{enumerate}
This structure enables regulators to verify grid-forming behavior 
independently of transient overload capability, 
reducing interpretive disputes during interconnection approval.

\subsubsection{Account for the voltage-divider effect at large phase jumps}
For $|\Delta\theta| \geq 30^\circ$, the geometric $P_{\text{poi}}$ 
drop from Proposition~1 becomes substantial: a $-60^\circ$ jump with 
representative parameters produces an immediate $\sim$15.5\% drop at 
the POI before any control action. At moderate-to-high pre-disturbance 
dispatch, this alone may bring $P_{\text{poi}}$ near or below any 
monotonic-rise threshold. Grid codes including large phase jumps should 
either exempt the initial sub-cycle transient from the monotonicity 
criterion or reduce the pre-disturbance dispatch for large-angle tests, 
ensuring that the criteria test the controller's response rather than 
the system's physical response.

\section{Conclusion}

Phase-jump performance criteria embed a quantitative hardware 
requirement, a minimum transient overload ratio, that depends on the 
test parameters ($\Delta\theta$, SCR, $X/R$, $P_{\text{pre}}$, $L_2$) 
but is nowhere stated in the specifications. This paper established a 
controller-architecture-independent physical bound on the achievable 
power-recovery trajectory by formulating the phase-jump response as a 
convex optimal control problem with a GFM voltage-tracking objective. 
The derived bound converts any phase-jump scenario into the corresponding 
minimum overload requirement. Validation against three WECC generic GFM models confirmed the bound 
is both valid and well-designed controller 
implementations can approach it closely, with the droop and hybrid 
models operating within 6--8\% of the theoretical frontier. 
The 34\% gap observed for the VSM model is attributable to the speed 
of the current-limiting mechanism rather than the synchronization 
philosophy, suggesting that direct, algebraic current enforcement 
offers a structural advantage over indirect, integral-based limiting 
under phase-jump conditions. For grid-code development, it is recommended that compliance evaluators distinguish between the inevitable geometric power excursion 
(Proposition~1), insufficient current margin, and genuine control 
deficiencies when interpreting test results. A two-tier test 
structure, such as reduced-dispatch for evaluating voltage-source fidelity and 
rated-dispatch for evaluating current-limited recovery, would 
separate these concerns. Future specifications should either state 
the minimum overload ratio explicitly or index performance criteria 
to the declared overload capability, eliminating the structural 
ambiguity that currently generates disputes during interconnection.
\bibliographystyle{IEEEtran}
\bibliography{gfmref}

\end{document}